\documentclass[aps,prl]{revtex4}
\usepackage{amsmath,amssymb}
\usepackage{color}
\usepackage[pdftex]{graphicx}

\begin{document}
\title{Vector Beam Bending}
\author{J. M. Nichols}
\affiliation{Naval Research Laboratory\\ 4555 Overlook Ave. SW.\\ Washington D.C. 20375}
\author{D. V. Nickel} 
\affiliation{Naval Research Laboratory\\ 4555 Overlook Ave. SW.\\ Washington D.C. 20375}
\author{F. Bucholtz}
\affiliation{Jacobs Technology, Inc.\\ 2551 Dulles View Drive\\ Herndon, VA 20171}

\begin{abstract}
It is well known that the trajectory of a light beam can curve if the medium in which the beam propagates is inhomogeneous or if the local spacetime is curved. In this paper we present a third possibility and show that a properly-prepared optical beam follows a curved trajectory even in free space where the spacetime is locally flat. By casting the problem in terms of transport of optical intensity, we show that a beam comprising a gradient in state of polarization to second order or higher (i.e., a specifically prepared vector beam) must also follow a curved trajectory. The effect is confirmed experimentally for the case of a second-order gradient in a linearly  polarized vector beam.
\end{abstract}
\maketitle


Optical beams with spatially varying polarization, i.e., ``vector beams'', have been found to possess useful properties that can be leveraged in application. Indeed, various methods for producing such beams have been developed \cite{Maurer:07},~\cite{Guo:17},~\cite{Fu:16} and subsequently applied to microscopy \cite{Bautista:17}, communications \cite{Qiao:17}, and material processing \cite{Orlov:18}. In this work we predict a new phenomenon associated with vector beams, namely a bending that results from spatial variations in the state of linear polarization. We forecast that such beams will ``self-bend'' upon exiting the aperture without requiring any variations in the properties of the medium through which the beam travels.  The physics described here differ fundamentally from Airy beam bending, where the intensity distribution shifts during propagation but the center of mass remains unchanged \cite{Siviloglou:07}. 
To arrive at our conclusion, we view beam propagation as a transport problem in which the spatial polarization derivatives are seen to produce a transverse ``force'' acting on the beam path.  The problem is solved in Lagrangian coordinates leading to the aforementioned predictions of curvature in the beam path.  We then verify these predictions in experiment by utilizing a liquid crystal spatial light modulator (SLM)-based optical setup to generate the required spatially-varying polarization across the beam.

For a monochromatic beam propagating in an isotropic, weakly inhomogeneous medium, the complex electric field amplitude vector is governed by
\begin{align}
    \nabla^2 {\bf E}(x,y,z)+ k_0^2n^2(x,y,z){\bf E}(x,y,z)=\{0,0,0\}.
    \label{eqn:waveqn}
\end{align}
where $k_0\equiv 2\pi/\lambda$, $\lambda$ is the wavelength of the light and $n(x,y,z)$ is the refractive index of the medium.  The beam properties in the transverse plane $X\equiv \{x,y\}$ are assumed to evolve slowly with respect to changes in the direction of propagation ($\hat{z}$) so that the assumed solution takes the transverse electromagnetic (TEM) form 
\begin{align}
    {\bf E}(x,y,z)=\rho(x,y,z)^{1/2}e^{i(k_0 z+\phi(x,y,z))}\left\{\cos(\gamma(x,y, z)),\sin(\gamma(x,y,z)),0\right\}.
    \label{eqn:sol}
\end{align}
In the model (\ref{eqn:sol}) we have 1) used a plane wave ``phasor'' representation with intensity $\rho$ and phase $\phi$ and 2) assumed a linear, fully polarized state defined by the angle $\gamma\equiv \tan^{-1}\sqrt{\rho^{(y)}/\rho^{(x)}}$, where $\rho^{(x)},~\rho^{(y)}$ are the projected intensities in the $\hat{x}$ and $\hat{y}$ directions.

Substituting (\ref{eqn:sol}) into (\ref{eqn:waveqn}) and collecting the imaginary portion of the result yields
\begin{align}
&\left(
\begin{array}{cc}
\frac{k_0\partial_z\rho+\nabla_X\cdot\left(\rho\nabla_X\phi\right)}{2\rho} & 0\\
0 & \frac{k_0\partial_z\rho+\nabla_X\cdot\left(\rho\nabla_X\phi\right)}{2\rho}
\end{array} 
\right) \left\{\begin{array}{c} \cos(\gamma)\\ \sin(\gamma) \end{array} \right\}
= 
\left(\begin{array}{cc}
k_0\partial_z\gamma+\nabla_X\gamma\cdot \nabla_X\phi & 0\\
0 & k_0\partial_z\gamma+\nabla_X\gamma\cdot \nabla_X\phi
\end{array} \right) \left\{\begin{array}{c} \sin(\gamma)\\ -\cos(\gamma) \end{array}\right\}
\label{eqn:contmat}
\end{align}
\normalsize
where we have invoked the paraxial assumption so that the gradient and Laplacian operators may be written with respect to the transverse coordinates $X$. Now define the ``velocity'' $\vec{v}(x,y,z)=k_o^{-1}\nabla_X\phi(x,y,z)$ as the change in optical path in the transverse direction per unit change in the direction of propagation (see e.g., \cite{Nichols:18}).  The only way for (\ref{eqn:contmat}) to be satisfied is if (a) $
\partial_z\rho+\nabla_X\cdot(\rho\vec{v})=0$ and (b) $
D\gamma/Dz=\partial_z\gamma+\nabla_X\gamma\cdot \vec{v}=0$ where the notation $D(\cdot)/Dz$ denotes the total derivative.  The first of these statements (a) is the well-known transport of intensity (TIE) equation (see e.g., \cite{Paganin:96},~\cite{Paganin:98}).
The second statement (b) holds that the polarization angle distribution across the face of the beam does not change with propagation distance (a known property of coherent beams \cite{Korotkova:05}).  We will leverage this property later in the derivation.

Gathering the real portion of (\ref{eqn:waveqn}), defining $\eta=(n^2-1)/2$, and again invoking the paraxial assumption yields
\small
\begin{align}
&\left(\begin{array}{cc}
-k_0^2\eta+k_0\partial_z\phi+\frac{1}{2}|\nabla_X\phi|^2+\frac{1}{2}|\nabla_X\gamma|^2-\frac{1}{2}\frac{\nabla_X^2\rho^{1/2}}{\rho^{1/2}} & \frac{\nabla_X\cdot (\rho\nabla_X\gamma)}{2\rho} \\
- \frac{\nabla_X\cdot (\rho\nabla_X\gamma)}{2\rho}  & -k_0^2\eta+k_0\partial_z\phi+\frac{1}{2}|\nabla_X\phi|^2+\frac{1}{2}|\nabla_X\gamma|^2-\frac{1}{2}\frac{\nabla_X^2\rho^{1/2}}{\rho^{1/2}} 
\end{array}
\right)\left\{\begin{array}{c} \cos(\gamma)\\ \sin(\gamma) \end{array} \right\}
=\left\{\begin{array}{c} 0\\ 0 \end{array} \right\}
 \label{eqn:eikonal1}
 \end{align}
 \normalsize
The solution requires that the matrix entries be zero, resulting in two equations.  Our focus is on the eikonal equation given by the main diagonal of (\ref{eqn:eikonal1}). 
Absent a polarization gradient, this term is exactly the eikonal of ref. \cite{Nichols:18} which predicts the beam path will be altered by the material index perturbations $\eta$ and where the final term involving $\rho$ predicts a diffractive beam broadening (see also ref. \cite{Nichols:19}). 
Taking the transverse gradient of the eikonal, ignoring diffraction, and invoking our prior definition of $\vec{v}$ yields
$D\vec{v}/Dz=-k_0^{-2}(\vec{\omega}\cdot\nabla_X)\vec{\omega}+\nabla_X\eta$
where $\vec{\omega}=\nabla_X\gamma$ is the transverse gradient of the polarization angle.  We thus arrive finally at the system of equations
\begin{subequations}
\begin{align}
    \partial_z\rho+\nabla_X\cdot(\rho\vec{v})&=0
    \label{eqn:continuity}
\end{align}
\begin{align}
    \frac{D\vec{\gamma}}{Dz}&=0
    \label{eqn:totalGamma}
\end{align}
\begin{align}
    \frac{D\vec{v}}{Dz}&=-\frac{1}{k_0^2}(\vec{\omega}\cdot\nabla_X)\vec{\omega}+\nabla_X\eta
\label{eqn:momentum}
\end{align}
\label{eqn:system}
\end{subequations}
\hspace*{-0.11in} which recasts our propagating TEM beam as a transport problem.  Importantly, by allowing a spatially varying polarization gradient we observe an additional ``forcing'' term in the eikonal governing the beam path (\ref{eqn:momentum}). The theory therefore predicts that through proper choice of spatial polarization distribution, one should be able to alter (bend) the path taken by the beam, even without changes in the properties of the medium (i.e., even with $\nabla_X\eta=0$).  Moreover, because the polarization angle distribution is independent of $z$ (Eqn. \ref{eqn:totalGamma}) the effect will persist over the the entire propagation distance and, by Eqn. (\ref{eqn:continuity}), will preserve the total beam intensity.

We now illustrate this effect experimentally. Consider the case where $\gamma$ varies only in $y$ and there are no index fluctuations. While the first component of the vector Eqn. (\ref{eqn:momentum}) states that the beam path will remain unchanged in $\hat{x}$, the second component becomes
\begin{align}
\frac{Dv}{Dz}&=-\frac{1}{k_0^2}\frac{d\gamma}{dy}\frac{d^2\gamma}{dy^2}.
\label{eqn:vel2}
\end{align}
Making the change to Lagrangian coordinates, the total derivative becomes an ordinary derivative and the transverse coordinates become functions of $z$, henceforth denoted $y\rightarrow y_z$ and $v_z=dy_z/dz$ (see e.g., \cite{Nichols:18}).  With this change, Eqn. (\ref{eqn:totalGamma}) becomes $d\gamma(y_z)/dz=0$ meaning that $\gamma(y_z)=\gamma(y_0)$ is a constant function of the initial (Eulerian) coordinates. Thus, Eqn. (\ref{eqn:vel2}) governing the optical beam path becomes
\begin{align}
\frac{d^2y_z}{dz^2}&=-\frac{1}{k_0^2} \frac{d\gamma(y_0)}{dy_0}\frac{d^2\gamma(y_0)}{dy_0^2}.
\label{eqn:accel2}
\end{align}
Solving (\ref{eqn:accel2}) by simply integrating w.r.t propagation distance $z$, and assuming $v_0=0$ (collimated light), yields the transverse displacement as a function of the polarization gradient
\begin{align}
y_z&=-\frac{z^2}{2k_0^2}\frac{d\gamma(y_0)}{dy_0}\frac{d^2\gamma(y_0)}{dy_0^2} +y_0.
\label{eqn:angledisp}
\end{align}
Defining the initial beam width $-a/2\le y_0\le a/2~m$, the largest polarization gradient we can observe is $\pi/a$ $\textrm{rad./m}$.  If we generate a beam with linear polarization profile  $\gamma(y_0)=\pi/2(y_0-a)^2/a^2-\pi/8$,
the transverse displacement (\ref{eqn:angledisp}) at a distance $z=L$ becomes (in meters)
\begin{align}
y_L&=\frac{L^2\pi^2}{2k_0^2a^4}\left(a-y_0\right)+y_0.
\label{eqn:path1}
\end{align}

\begin{figure}[th]
  \centerline{
   \begin{tabular}{ccc}
    \includegraphics[scale=0.53]{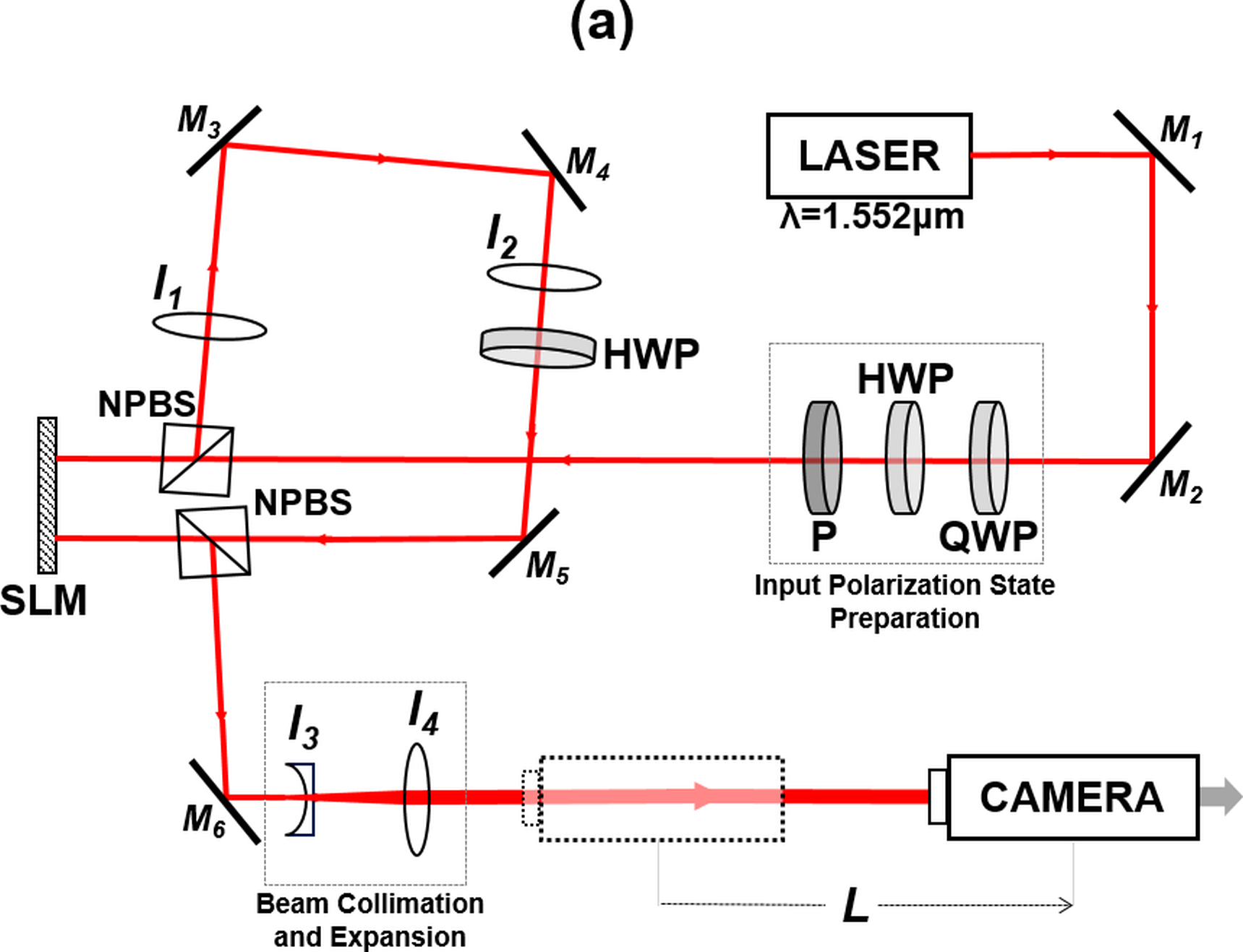} & \hspace*{0.0in} & 
    \includegraphics[scale=0.55]{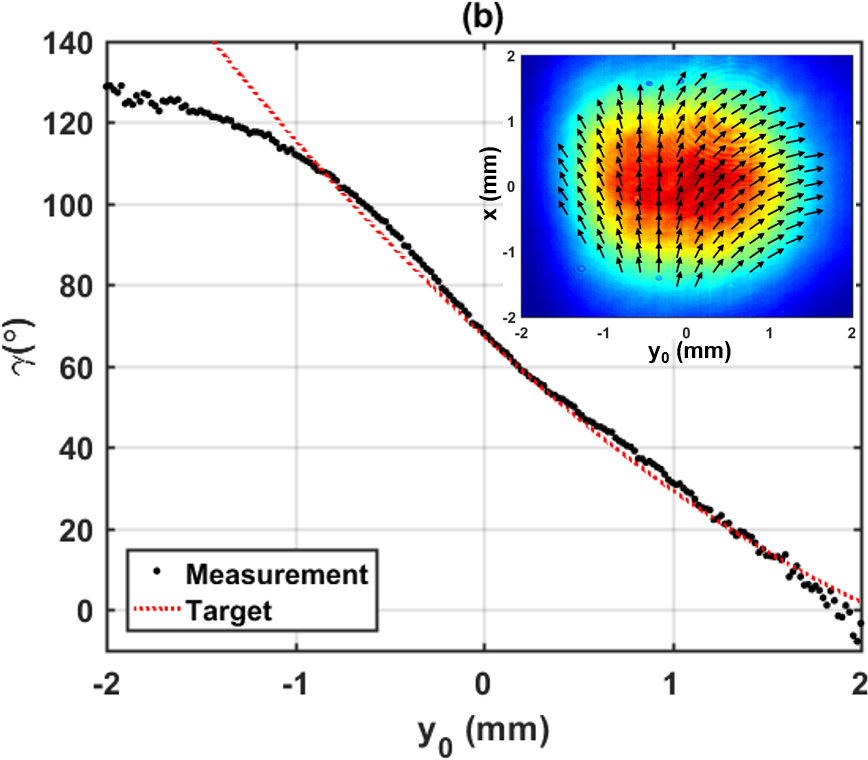} 
  \end{tabular}
  }
  \caption{(a) Diagram of the optical setup adapted from ref. \cite{Runyon:18} used to generate the desired $\gamma(y_0)$. (b) Targeted (red line) and measured $\gamma(y_0)$ across the center of the beam face (sliced at $x=0~mm$) (black points) . The inset shows the full measured 2-D polarization angle distribution of the beam (arrows) overlaid upon its normalized intensity profile. M: mirror, QWP: quarter-wave plate, HWP: half-wave plate, P: linear polarizer, NPBS: non-polarizing beam-splitter, l: lens.}
  \label{fig:setup}
\end{figure}

To test (\ref{eqn:path1}) we used the optical setup shown in Figure \ref{fig:setup} (a) along with the theory and methods described in \cite{Runyon:18} and \cite{sit:17}.  Specifically, a $\lambda=1.552~\mu m$ beam was modified by an SLM with spatial dimension $a=4.17~mm$ to generate the desired polarization profile $\gamma(y_0)$.  Figure (\ref{fig:setup}) (b) shows the measured polarization angle across the center (sliced at $x=0~mm$), the ideal targeted polarization angle distribution, and the full measured 2-D polarization angle distribution of the beam overlaid upon its normalized intensity profile (Fig. \ref{fig:setup} (b) inset). The beam was almost completely linearly polarized with an eccentricity of $>0.98$ across the entire beam face. The transverse displacement was measured as a function of distance by capturing the profiles of the unperturbed (SLM phase masks off) and perturbed (SLM phase masks on) beam every $0.5m$ using an infrared camera mounted on a movable cart. To account for beam pointing errors due to turbulence, multiple beam profiles were captured at each position in order to determine the mean relative displacements. 

\begin{figure}[th]
  \centerline{
   \begin{tabular}{ccc}
    \includegraphics[scale=0.63]{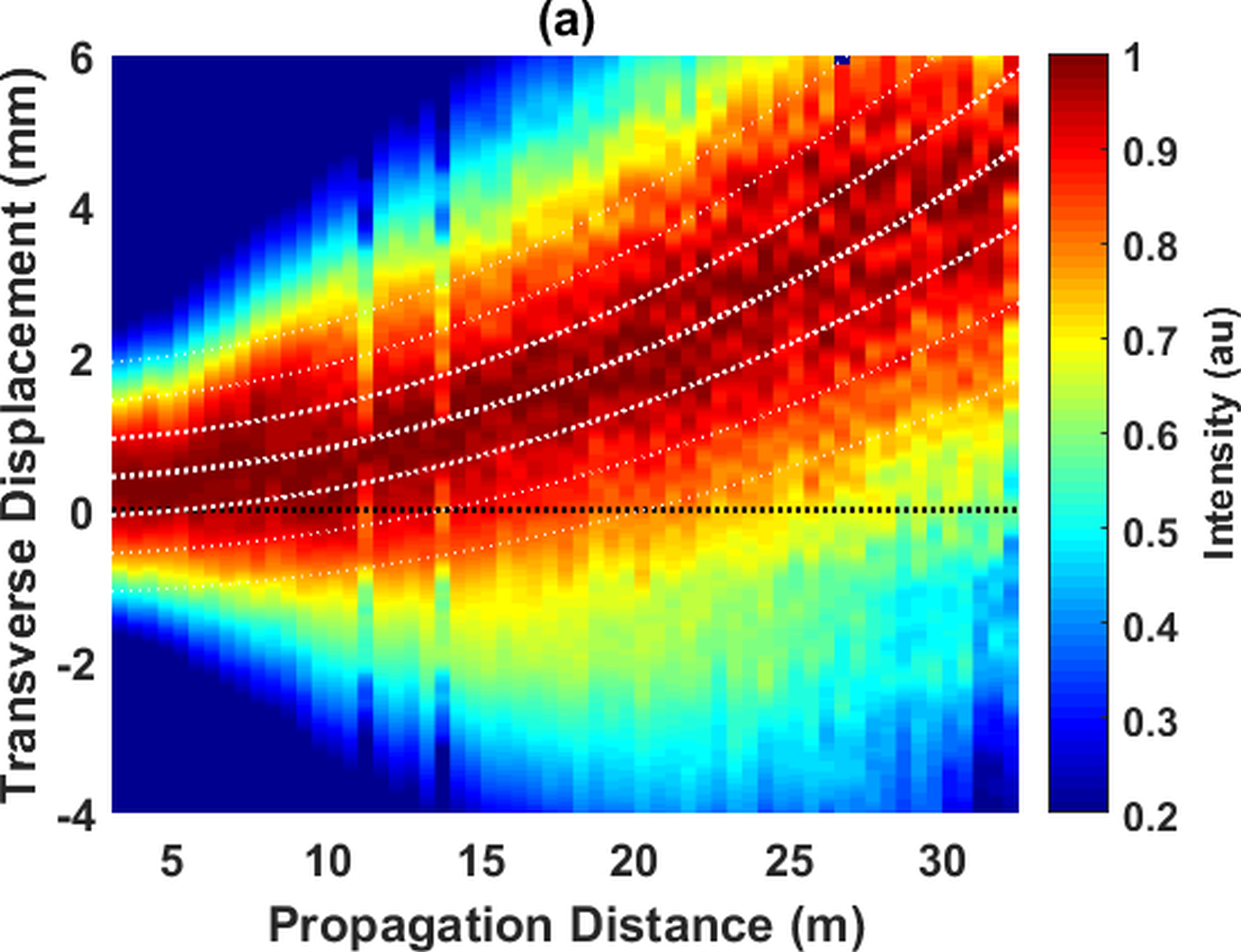} & \hspace*{0.0in} & 
    \includegraphics[scale=0.5]{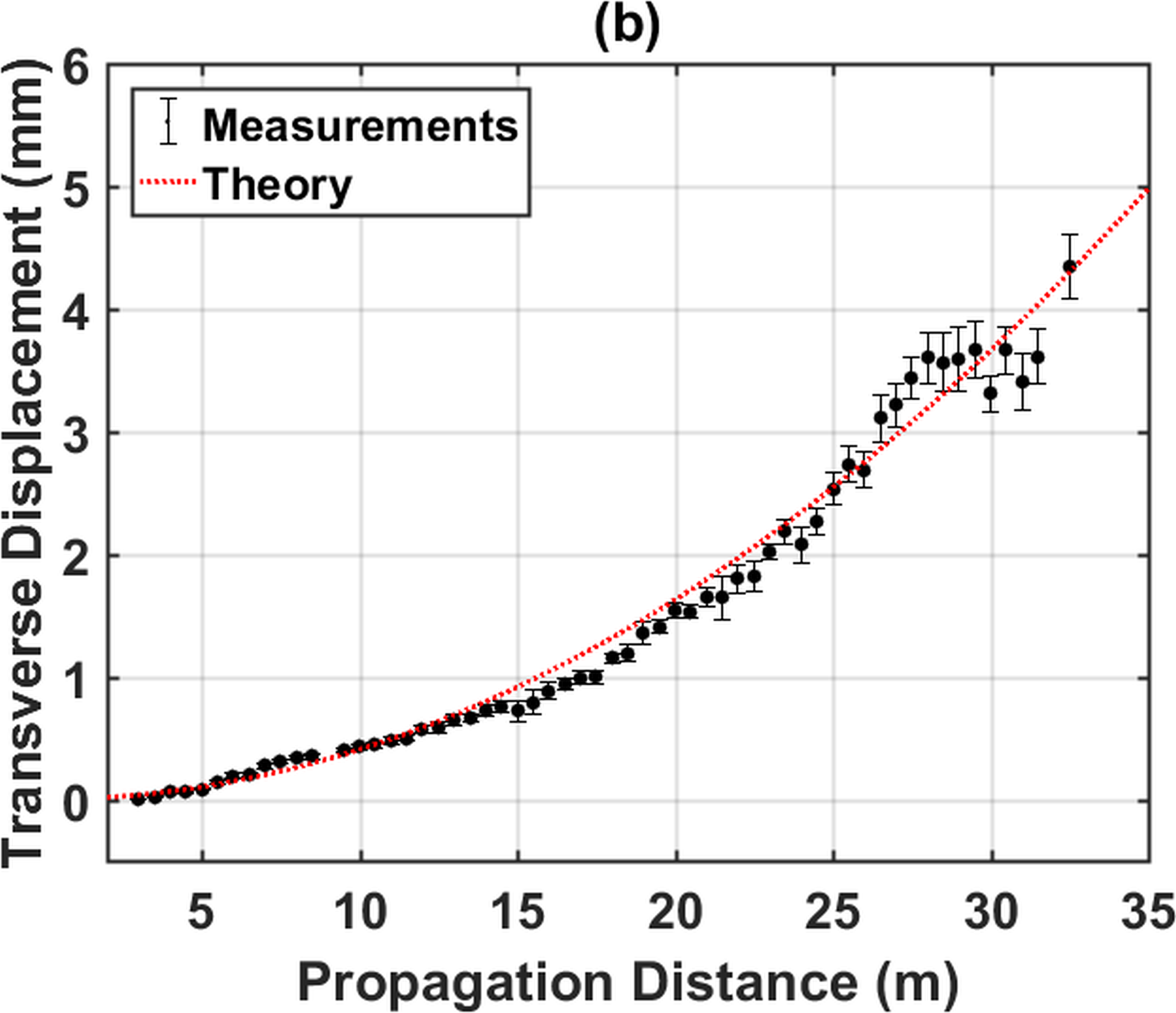}
  \end{tabular}
  }
  \caption{(a) Normalized y-axis intensity slices taken from single images of the beam vs. propagation distance . Also plotted are the paths predicted using Eqn. (\ref{eqn:path1}) for various values of $y_0$ and $a=4.17~mm$ (white dashed lines). (b) The mean transverse displacement of the beam's maximum intensity peak (black data points) and the curve predicted using Eqn. \ref{eqn:path1} with $y_0$=0 and $a=4.17~mm$ (red dashed line) plotted as a function of distance.}
  \label{fig:res1}
\end{figure}

Figure (\ref{fig:res1}) (a) shows the normalized intensity profiles of the generated beam plotted as a function of propagation distance as well as the paths predicted using Eqn. (\ref{eqn:path1}) for various values of $y_0$. Figure (\ref{fig:res1}) (b) shows the mean transverse displacement of the beam's maximum intensity peak determined from the 1200+ captured beam profiles (10 each every $0.5m$). The measured transverse displacements agree extremely well with the curve predicted using Eqn. (\ref{eqn:path1}) with $y_0=0~mm$ and $a=4.17~mm$.
While our focus was on predicting the beam path, we note that the full model (\ref{eqn:eikonal1}) does include the standard diffractive beam broadening.  For a Gaussian beam model, solving (\ref{eqn:accel2}) with the diffractive term included yields a beam profile matching that shown in Fig. (\ref{fig:res1}a).  While diffraction has no substantive influence on the beam curvature explored in this work, it is necessary to accurately forecast beam width during propagation.  

In this paper we have predicted an effect whereby a vector beam can self-bend in a manner that depends on the spatial distribution of the linear polarization angle.  Using Lagrangian coordinates in conjunction with a ``transport'' formulation of optical propagation dynamics, the degree of bending can be predicted analytically as a function of propagation distance.  We subsequently demonstrated the accuracy of these predictions in experiment.  While the magnitude of the effect is small (scales as $k_0^{-2}$) it is certainly measurable, particularly over long distances, and provides a non-mechanical approach to beam steering.  We also predict that other states of polarization may yield still other, qualitatively different beam trajectories.  More generally, a new mechanism for curving an optical beam in a locally flat spacetime may have implications in other areas of physics. For example, the effect may play a role at cosmological scales where recent evidence \cite{M87} indicates that millimeter-wave radiation from black holes exhibits both partial polarization and polarization gradients.
\bibliographystyle{plain}
\bibliography{refs}
\end{document}